# FEATURE SELECTION ON BOOLEAN SYMBOLIC

Djamal Ziani[1]

[1]College of Computer and Information Systems Sciences, Information Systems Department, PO Box 51178 Riyadh 11543 Saudi Arabia
dziani@ksu.edu.sa

## ABSTRACT

*With the boom in IT technology, the data sets used in application are more and more larger and are described by a huge number of attributes, therefore, the feature selection become an important discipline in Knowledge discovery and data mining, allowing the experts to select the most relevant features to improve the quality of their studies and to reduce the time processing of their algorithm. In addition to that, the data used by the applications become richer. They are now represented by a set of complex and structured objects, instead of simple numerical matrixes. The purpose of our algorithm is to do feature selection on rich data, called Boolean Symbolic Objects (BSOs). These objects are described by multivalued features. The BSOs are considered as higher level units which can model complex data, such as cluster of individuals, aggregated data or taxonomies. In this paper we will introduce a new feature selection criterion for BSOs, and we will explain how we improved its complexity.*

## 1. INTRODUCTION

To facilitate discrimination between objects the experts use feature selection algorithms in order to select a subset of features which are the most discriminant, without deteriorating the reliability of the data. The problem of feature selection has often been treated in classical data analysis (e.g., discriminant analysis); many techniques or algorithms have been proposed [1].

In classical data analysis, the data are represented by an array of individual × variables, and the goal of discrimination is to distinguish between the classes of individuals. With the growth of databases, it becomes very important to summarize these data by using a complex type called symbolic object [2] [3].

The feature selection algorithms on symbolic objects need complex calculation, since the data processed may represent classes of real individuals and each variable is not limited to one value, but may indicate a distribution of values [4]. Therefore our study will focus on two main elements: choosing of a good dissimilarity measure to select powerful variables, and improving the algorithm complexity.

The purpose of the feature selection algorithm which we present here is to find the minimum set of variables which contribute in the discrimination between different symbolic objects, and eliminate the variables which not contribute in discrimination or which contribution is already covered by the selected variables.

Since symbolic objects are complex data, very few feature selection algorithms have been developed until now in Symbolic Data Analysis. The selection criteria used in these algorithms must deal with many types of data (alphabetic, numeric, intervals, set of values), and should measure a dissimilarity between symbolic objects. Also these algorithms should be well optimized to support complex calculations. In our paper we will present a new feature selection algorithm with a strong selection criterion, which can deal with any type of data; and we will see how this algorithm will be optimize in order to process feature selection on large symbolic objects datasets.

## 2. SYMBOLIC OBJECTS

Before we present the algorithm, we will give some definitions of symbolic objects [5]:
$Y= \{y_1, ..., y_n\}$ is the set of variables;
$O= \{O_1, ..., O_n\}$ is the set of domains where each variable takes its values;
$\Omega=\{w_1, ..., w_p\}$ is the set of elementary observed objects;
$\Omega' = O_1 \times .... \times O_n$ is the set of all possible elementary observed objects.
An elementary event is represented by the symbolic expression $e_i = [y_i=v_i]$, where $v_i \subset O_i$, and it is defined by $e_i: \Omega \rightarrow \{true, false\}$ as $e_i(w)= true \rightarrow y_i(w) \in v_i$.
An assertion object is a conjunction of elementary events, represented by the symbolic expression: $a = [y_1=v_1] \wedge ... \wedge [y_n=v_n]$.

### Example 1:

Let us have two elementary observed objects. Suppose *Alain* has brown hair and *John* has grey hair. If "hair" is considered as a variable, this means that *hair(Alain) = brown and hair(John) = grey*.

The elementary event $e_1 = [hair = \{brown, black\}]$ is such that:
$e_1$(Alain) = true, since hair(Alain)=brown $\in$ {brown, black}.

If Alain is tall and John is short we can build the following assertion: a = [hair ={brown, black}] $\wedge$ [height ={tall, small}]

We distinguish two types of extents:
- The real extent of the symbolic object *a* is defined referring to $\Omega$, and present the set of elementary objects observed which satisfy: $ext_\Omega(a)=\{w_l \in \Omega | a(w_l) = true\}$.

- The virtual extent of symbolic object *a* is defined referring to $\Omega'$, and present the set of elementary objects observed which satisfy: $ext_{\Omega'}(a) = \{w_l' \in \Omega' | a(w_l') = true, i.e., \forall y_i \; y_i(w_l') \in v_i \; and \; v_i \in V_i\}$, where $v_i$ is a value taken by the variable $y_i$ in the object $w_l'$ and $V_i$ is a value taken by the variable $y_i$ in the assertion *a*.

### Example 2:

Let $\Omega$={Alain, John, Sam}
hair(Alain) = brown and height(Alain) = tall
hair(John) = grey and height(John) = tall
hair(Sam) = black and height(Sam) = small.
a = [hair ={brown, black}] $\wedge$ [height ={tall, small}]
$ext_\Omega$(a) = {Alain, Sam}.
Then, $ext_{\Omega'}$(a) = {Alain, Sam, all virtual individuals with brown or black hair, and tall or small height}.

### 2.1. Notion of discrimination

Since our feature selection algorithm criteria are based on discrimination, let us explain the discrimination on symbolic objects through the following symbolic objects:

$a_1$=[ age = [25,45]] $\wedge$ [weight = [65,80[ ]
$a_2$=[ age = [15,35]] $\wedge$ [weight = ]80,90] ]
$a_3$=[ age = [20,35]] $\wedge$ [weight = [70,85] ].

There are two types of discrimination, Boolean and Partial:

The boolean or total discrimination between symbolic objects means that there is an empty intersection between the virtual object extents. This is what we have with the objects $a_2$ and the objects $a_3$. See Figure 1. The discrimination between these two is equal to 1 (maximum).

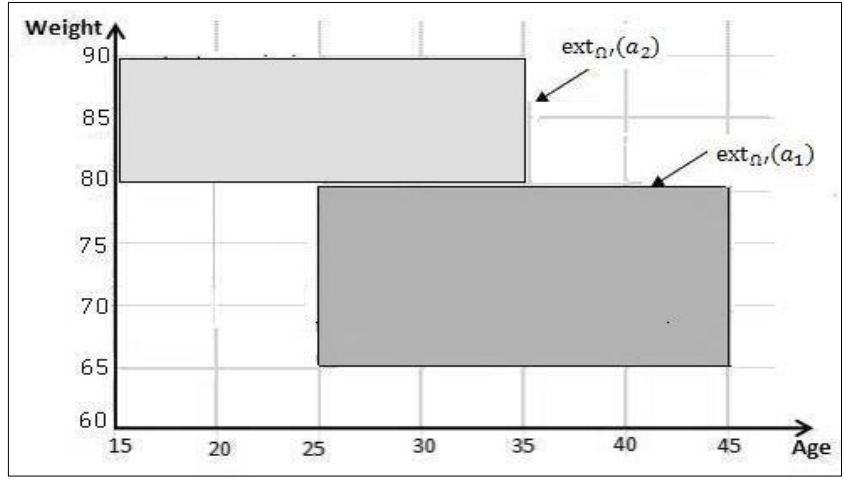

Figure 1. Total discrimination

The partial discrimination between symbolic objects means that there is an intersection between the virtual object extents. This is what we have with the objects $a_1$ and the objects $a_3$. See Figure 2.

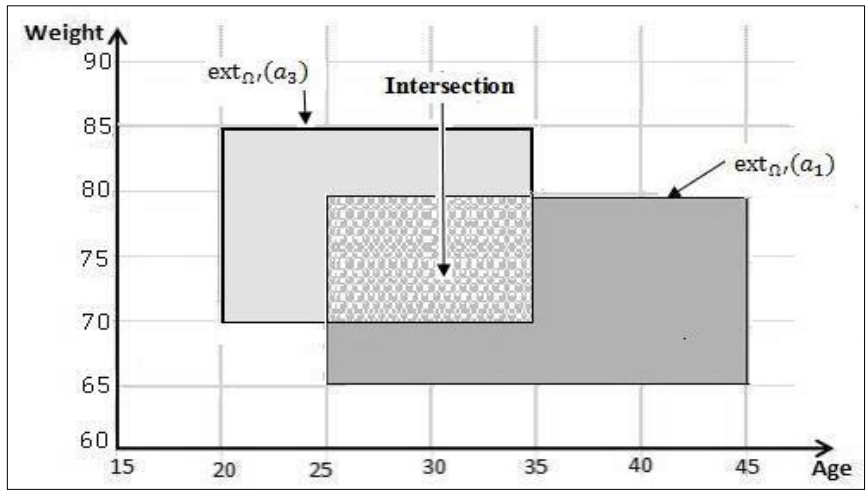

Figure 2. Partial discrimination

### 2.2. Dissimilarity measures

Dissimilarity measures are used as a basis for selection criteria in feature selection. In the literature we can find many dissimilarity measures, but only a few can be considered when working with symbolic objects data. In this section we will consider some of the dissimilarity measures used for symbolic objects.

We will assess the dissimilarity measures base on the following criteria:
Mathematical properties: reflexive (an object is similar to itself), and symmetric (if $s$ is a similarity measure $s(a_i,a_j)=s(a_j,a_i)$)

- Type of feature: qualitative, quantitative
- Boolean or partial discrimination
- Complexity of the calculation.

We will use the same notation for all measures.

Let $v_{il}$, $v_{jl}$ be the values taken by the variable $y_l$ in the assertion $a_i$ and $a_j$.

Let $\mu(.)$ compute either the length, if the feature is of an interval type, or the number of elements included in a set for categorical type features. Also,

$\oplus$: computes the overlapping of area.

$\otimes$: Cartesian product which computes the union area.

- Vignes indice of dissimilarity [6]

$$comp(v_{il}, v_{jl}) = \begin{cases} 1 \text{ if } v_{il} \cap v_{jl} = \emptyset, \\ 0 \text{ else.} \end{cases} \quad (1)$$

This indice is reflexive and symmetric, it can be used with quantitative and qualitative features, it is based on boolean discrimination, and it is not complex since it is using only an intersection operator.

- Kiranagi and Guru's dissimilarity measure [7]

$$d(v_{il}, v_{jl}) = \frac{\mu(v_{il}) + \left(max\left(f_{il}^{-1}, f_{jl}^{-1}\right) - min\left(f_{il}^{+1}, f_{jl}^{+1}\right)\right)}{\mu(v_{jl})} \quad (2)$$

Where $v_{il} = [f_{il}^{-1}, f_{il}^{+1}]$ and $v_{jl} = [f_{jl}^{-1}, f_{jl}^{+1}]$.

This measure is not reflexive and not symmetric; therefore, it is not good for our algorithm.

- De Carvalho's dissimilarity measure based on potential description [8]

$$d(v_{il}, v_{jl}) = 1 - \frac{\mu(v_{il} \oplus v_{jl})}{\mu(v_{il} \oplus v_{jl}) + \mu(v_{il} \oplus \overline{v_{jl}}) + \mu(\overline{v_{il}} \oplus v_{jl})} \quad (3)$$

where $\overline{v_{il}}$: is the complement of $v_{il}$.

This dissimilarity measure is reflexive and symmetric, it can be used with quantitative and qualitative features, it calculates partial discrimination, and it is complex since it calculates two times the complementary values of a variable, and this operation is complex.

- Ichino and Yaguchi's first formulation of a dissimilarity measure [9]

$$\phi(v_{il}, v_{jl}) = \mu(v_{il} \oplus v_{jl}) - \mu(v_{il} \otimes v_{jl}) + \gamma\left(2\mu(v_{il} \otimes v_{jl}) - \mu(v_{il}) - \mu(v_{jl})\right) \quad (4)$$

where γ is as an arbitrary value in the interval [0, 0.5].

This dissimilarity measure is reflexive and symmetric, it can be used with quantitative and qualitative features, it is based on partial discrimination, and it is complex since it is using two times the Cartesian product $\otimes$.

- Dissimilarity indice of Ravi and Gowda [10]

Qualitative feature: $D(v_{il}, v_{jl}) = D_s(v_{il}, v_{jl}) + D_c(v_{il}, v_{jl})$ (5)

where $D_s(v_{il}, v_{jl}) = cos(90 * (\mu(v_{il}) + \mu(v_{jl}))/2l_s)$
$l_s = \mu(v_{il}) + \mu(v_{jl}) - \mu(v_{il} \cap v_{jl})$, and $D_c(v_{il}, v_{jl}) = cos(90 * \mu(v_{il} \cap v_{jl})/2l_s)$.

Quantitative feature: $D(v_{il}, v_{jl}) = D_p(v_{il}, v_{jl}) + D_s(v_{il}, v_{jl})$ (6)

where $D_p(v_{il}, v_{jl}) = \cos\left(90 * \left(1 - (\mu(v_{il}) - \mu(v_{jl}))/\mu(O_l)\right)\right)$ and $D_c(v_{il}, v_{jl}) = \cos\left(90 * (\mu(v_{il}) + \mu(v_{jl})/2l_s)\right)$.

This dissimilarity measure is symmetric but not reflexive, it can be used with quantitative and qualitative features, it is based on partial discrimination, and it more complex than (1) and (2), but less complex than (3) and (4).

Based on the remarks we have listed for the shown measures, we choose Vignes indice of dissimilarity as the basis of our feature selection criterion, but we will change it in order to be able to calculate partial discrimination.

## 3. SELECTION CRITERIA

### 3.1. Discrimination Power

Let *A* be a set of assertions, *n* is the number of assertions in *A*, *Y* is a set of variables, *K* is the set of assertion pairs *K=AxA*, and *P(Y)* is the set of subsets of *Y*. The function *comp* used in (1) indicates the existence or not of an intersection between two objects. We say that two assertions $a_i$ and $a_j$ are discriminated by a variable *y* if and only if *comp($v_{il}$, $v_{jl}$) =1*. The discriminant power of a variable $y_l$ on the set *K*, denoted by *DP($y_l$, K)*, is equal to the number of assertion pairs discriminated by the variable $y_l$: i.e.,

$DP: Y \times K \rightarrow \mathbb{N}$ where $\mathbb{N}$ is the set of integers

$$DP(y_l, K) = \sum_{i=1}^{n-1} \sum_{j=i+1}^{n} comp(v_{il}, v_{jl}) \; with \; (a_i, a_j) \in K. \tag{7}$$

The discriminant power of a variable subset *Yd* is equal to the number of assertion pairs discriminated at least by one variable of *Yd*: i.e.,

$DP: P(Y) \times K \rightarrow \mathbb{N}$

$$DP(Yd, K) = \sum_{i=1}^{n-1} \sum_{j=i+1}^{n} max_{y_l \in Yd} \; comp(v_{il}, v_{jl}). \tag{8}$$

**Example 3:**

Let us have the following 3 assertions:
$a_1$ = [ height = tall ] ∧ [ hair = brown ]
$a_2$ = [ height = {tall, medium} ] ∧ [ hair = black ]
$a_3$ = [ height = small ] ∧ [ hair = {brown, black} ];
DP(height, K) = card({$a_1$, $a_3$), ($a_2$, $a_3$)} = 2
DP(hair, K) = card({$a_1$, $a_2$)}) = 1
DP({height, hair}, K) = card({($a_1$, $a_2$), ($a_1$, $a_3$), ($a_2$, $a_3$)}) = 3

### 3.2. Original Discrimination Power

The original discrimination power, denoted *ODP*, of a variable $y_l$ referred to a set of variable *Yp*, is equal to the number of assertion pairs discriminated by $y_l$ and not discriminated by any variable of *Yp*. That is,

$ODP: Y \times P(Y) \times K \rightarrow \mathbb{N}$
$$ODP(y_l, Yp, K) = \sum_{i=1}^{n-1} \sum_{j=i+1}^{n} max\left(comp(v_{il}, v_{jl}) - max_{y_p \in Yp}\left(comp(v_{ip}, v_{jp})\right), 0\right), with \; (a_i, a_j) \in K.. \tag{9}$$

The external *Max*, in the expression, constrains the function to have null or a positive value for each assertions pair.

**Example 4:**

Using the assertions $a_1$, $a_2$, and $a_3$ of the example 3, we obtain: *ODP(height, hair, K)=2*.

## 4. MINSET ALGORITHM

In the literature, we can find only very few algorithms which can treat feature selection for symbolic objects. We can cite, the algorithm of Ichino [11] uses geometrical thickness criterion to select the features, the algorithm of Nagabhushan et al. [12] and Kiranagi et al. [13] select features for clustering objects using similarity between symbolic objects, and the algorithm of Chouakria et al. [14] is for principal components. In our paper, we will concentrate on the algorithm of Vignes [6] called *Minset*, since it is the only algorithm doing feature selection on symbolic object with the purpose to discriminate between objects. And we will propose in this paper a new algorithm called *Minset-Plus* base on the improvement made on *Minset* algorithm.

To discriminate the objects, the *Minset* algorithm proceeds by finding discrimination between all assertion pairs. The underlying idea of this treatment results from the fact the discrimination between symbolic objects is based on the intersection of their virtual extensions. The idea of the *Minset* algorithm is formalized as follows. We have as entry the knowledge base *(Y,O, A)*. The objective of the algorithm is to find another knowledge base *(Y',O',A)* such that $Y' \subseteq Y$ with $DP(Y', K) = DP(Y, K)$. The subsets *Y* and *Y'* represent two sets of variables. The algorithm has three principal steps:

The first step is to select the indispensable variables. A variable is considered as indispensable, if when you remove it from the set of variables, the discrimination power of this set of variables will be reduce (less than the discrimination power of all variables)

The second step in the algorithm is to select in each iteration a variable which has the greater value in discriminating the symbolic objects, not discriminate yet by the selected variables.

The third step will illuminate a variable which become redundant, it means that the part of discrimination bring by this variable, have been covered by a combination of the other selected variables.

1. Find the indispensible variables which permit to discriminate assertion pairs not discriminated by other variables. This means we select all other variables such that their ODP against all other variables is ≠0: ODP(yi,Y-yi, K)≠0.
Set Y' = Y
Set $Yd = set\ of\ selected\ variables$
While DP($Yd$, K) < DP(Y, K)
2. Select in each step the variable which has the highest ODP. The selected variable permits to discriminate the greatest number of assertion pairs, not already discriminated by the variables selected before.
Y'= Y' - Selected variables
$Yd = Yd \cup \{ y_l\ /\ y_l\ maximizes\ ODP(y_i, Y' - y_i, K)\ \forall y_i \in Y'\}$
3. Eliminate in each step the variables which become redundant. This means the assertion pairs discriminated by these variables are discriminated by other selected variables.
$Yd = Yd - \{y_l \in Yd\ where\ ODP(y_l, Yd - y_l, K) = 0\}$

## 5 MINSET-PLUS ALGORITHM

### 5.1. Critics of Minset Algorithm

To calculate the $ODP(y_l, Y_p, K)$, the algorithm must execute k×(1+p) times the function comp where *k= card(K)* and *p = card(Yp)*. The $DP(y_l, K)$ is calculated with k executions of the function comp. The discovery of some mathematical properties relating to the function *ODP* and *DP* has permitted us to reduce considerably the temporal complexity of the algorithm (see section 6).

The object class's outlines are often fuzzy because they depend on the subjectivity of the experts; so, the boolean discrimination between object classes is a strong hypothesis, which may deteriorate the finality of the study. In this case, we must introduce a new comparison function, which can be founded on a partial discrimination between symbolic objects.

### 5.2. Definition of the new discrimination function

The new discrimination function between symbolic objects will be calculated by the function g. This function can calculate both boolean and partial discrimination [15]. This partial discrimination notion has been introduced in the algorithm *Minset-Plus* in order in order to avoid neglecting that information part provided by the variables when there is no total discrimination. The function *g* which calculates partial discrimination is defined as follows:

$$g: K \times Y \to [0,1]$$
$$g\left((a_i, a_j), y_l\right) = 1 - \frac{card(v_{il} \cap v_{jl})}{card(v_{il} \cup v_{jl})} \tag{10}$$

### 5.3. Use of mathematical properties to reduce complexity

Before we demonstrate some useful mathematical properties of the *ODP* and *DP* functions, let us give some definitions:

$$DP(y_l, K) = \sum_{i=1}^{n-1} \sum_{j=i+1}^{n} g\left((a_i, a_j), y_l\right), with\ (a_i, a_j) \in K \tag{11}$$

$$DP(Yd, K) = \sum_{i=1}^{n-1} \sum_{j=i+1}^{n} max_{y_p \in Yd}\ g\left((a_i, a_j), y_p\right) \tag{12}$$
with $(a_i, a_j) \in K\ and\ Yd \subseteq Y$

$$ODP(y_l, Yp, K) = \sum_{i=1}^{n-1} \sum_{j=i+1}^{n} max\left(g\left((a_i, a_j), y_l\right) - max_{y_p \in Yp}\left(g\left((a_i, a_j), y_p\right)\right), 0\right), \tag{13}$$
with $(a_i, a_j) \in K$.

### Property 1

The following property permits us to calculate the discrimination power of a set of variables using the selected power of the old selected variable and the original discrimination power of the current selected variable. The benefit of this is to not calculate in each step the discrimination power of the selected variable.

$$DP(Y_P \cup y_i, K) = DP(Y_P, K) + ODP(y_i, Y_P, K). \tag{14}$$

### 5.4. Discrimination matrix

We can see that the calculation of the *DP* and *ODP* functions is based on the calculation of $g((a_i, a_j), y_l)$. This is done repetitively in each step. Furthermore, we know that the g function involves ∪ and ∩ operations between a set of values, and these operations are not easy operations. Therefore, this is why we need to find a way to avoid calculating the same thing many times. We will save the old calculations to reuse them in further algorithm steps. This idea can be done by the introduction of a discrimination matrix.

This discrimination matrix allows us to calculate only one time $g((a_i, a_j), y_l)$, and during all the steps of the algorithm we will use the matrix to do all the necessary operations. This is a huge optimization of the time complexity. Also, the size of this matrix is not big, it is: k ×

$n$ where $k = card(K)$ and $n = card(Y)$, $K$ is not a big number, since we are dealing with classes of individuals.

**Example 5:**

Let us have $Y= \{y_1, y_2, y_3, y_4, y_5\}$ and $A = \{a_1, a_2, a_3, a_4\}$. At the beginning the algorithm must calculate the stopping criterion (*While DP(Yd, K) < DP(Y, K)*). *DP(Y, K)* is the total Discrimination power of all variables of the dataset: *DP(Y, K)*. By using (12):

$$DP(Y,K) = \sum_{i \neq j}^{n} \sum_{j \neq i}^{n} max_{y_l \in Y} \, g\left((a_i, a_j), y_l\right)$$

We can notice that to calculate *DP(Y, K)*, we compute all $g((a_i, a_j), y_l)$, and this is done many times during the algorithm execution, so to avoid repetitive and unnecessary calculations, we decide to save, during the calculation of the stopping criterion, all $g((a_i, a_j), y_l)$ values in the discrimination matrix. Each $g((a_i, a_j), y_l)$ will be saved in the matrix case corresponding to row $y_l$ and the column $(a_i, a_j)$.

Also we will calculate for each pair of object $(a_i, a_j)$, the maximum discrimination obtained by the variables. This operation is done by filling *Max Yd* row.

*Max Yd* row is used to save: $max_{y_d \in Yd} \, g\left((a_i, a_j), y_d\right)$.

At the beginning *Max Yd* is empty, and then we put the maximum value $g((a_i, a_j), y_p)$ of the indispensible variables $y_p$ is an indispensible variable). Here, in Table 1, $y_1$ is indispensible.

Table 1. Discrimination Matrix

|        | $(a_1, a_2)$ | $(a_1, a_3)$ | $(a_1, a_4)$ | $(a_2, a_3)$ | $(a_2, a_4)$ | $(a_3, a_4)$ |
|--------|---|---|---|---|---|---|
| $y_1$  | 0.7 | 0   | 0.1 | 0.1 | 0   | 0.1 |
| $y_2$  | 0   | 0.6 | 0.1 | 0.7 | 0   | 0.5 |
| $y_3$  | 0   | 0.3 | 0.5 | 0.3 | 0.6 | 0.1 |
| $y_4$  | 0   | 0.6 | 0.1 | 0.7 | 0.6 | 0.5 |
| $y_5$  | 0   | 0.3 | 0.1 | 0.3 | 0.2 | 0.3 |
| Max Yd | 0.7 | 0.6 | 0.5 | 0.7 | 0.6 | 0.5 |

### 5.5. Algorithm operations using discrimination matrix

To select the variables, the *Minset-Plus* algorithm will use the same steps as seen in section 4. However, based on the discrimination matrix, the operations used in each step will be different and less complex. To select a new variable in each step, we have to calculate for each none selected variable $ODP(y_l, Yd, K)$. We know:

$$ODP(y_l, Yd, K) = \sum_{i=1}^{n-1} \sum_{j=i+1}^{n} max\left(g\left((a_i, a_j), y_l\right) - max_{y_p \in Yd}\left(g\left((a_i, a_j), y_p\right)\right), 0\right)$$

We know that the value of $g\left((a_i, a_j), y_l\right)$ is saved in the discrimination matrix, and exactly in the case corresponding to the row $y_1$ and column $(a_i, a_j)$. Since $max_{y_p \in Yd}\left(g\left((a_i, a_j), y_p\right)\right)$ calculates the maximum value of $g\left((a_i, a_j), y_p\right)$ for a set of variables, this value will be saved in the case corresponding to the column $(a_i, a_j)$. and the row *Max Yd*. Thus, the calculation of $ODP(y_l, Yd, K)$ will take only one subtraction of numbers and one comparison to find the maximum between 2 numbers.

In order to find the redundant variables in each step, we will have only to calculate for each pair $(a_i, a_j)$, the subtraction of the value of the case corresponding to the column $(a_i, a_j)$ and row $y_l$ and the value of the case corresponding to the row $y_s$ and column $(a_i, a_j)$, and then we compare the value of the calculated subtraction with the value of the case corresponding to the row *Max Yd* of the same column *($a_i$, $a_j$)*.

The selection of indispensible variables is now done by using partial discrimination; therefore, the formulation of the indispensible variables will change a little bit. We can say that a variable is indispensible if we find a pair of objects discriminated totally or partially by the variable and not discriminated at all by any other variable:

$y_l$ is indispensible if:

$$\exists (a_i, a_j) \in K \text{ where } g((a_i, a_j), y_l) \neq 0 \text{ and } g((a_i, a_j), y_l) - \max_{y_p \in Y - y_l}(g((a_i, a_j), y_p)) > 0. \tag{15}$$

**Example 6:**

Using the discrimination matrix of example 3, we have the following algorithm steps:

- Calculation of stopping criteria: the power discrimination of the all variable is calculated by doing the sum of all values in the row Max Yd.

$$DP(((y_1, y_2, y_3, y_4, y_5)), K) = 3.6$$

- Finding indispensable variables.

By using the discrimination matrix of example 3, we can deduct that y1 and y3 are indispensables since:

$$g((a_1, a_1), y_1) = 0.7 \neq 0 \quad and \quad g((a_i, a_j), y_l) - \max_{y_p \in Y - y_1}(g((a_i, a_j), y_p)) = 0.7 > 0.$$

$$g((a_1, a_1), y_3) = 0.5 \neq 0 \quad and \quad g((a_i, a_j), y_l) - \max_{y_p \in Y - y_3}(g((a_i, a_j), y_p)) = 0.1 > 0.$$

We update *MaxYd* with the maximum discrimination values reached by the indispensable variables (see Table 2). By doing the summation of the values of Max Yd we get the discrimination power of the indispensable variables:

$$DP(((y_1, y_3)), K) = 2.5$$

Table 2. Max Yd updated with value of indispensable variables

|        | ($a_1$, $a_2$) | ($a_1$, $a_3$) | ($a_1$, $a_4$) | ($a_2$, $a_3$) | ($a_2$, $a_4$) | ($a_3$, $a_4$) |
|--------|------|------|------|------|------|------|
| $y_1$  | 0.7  | 0    | 0.1  | 0.1  | 0    | 0.1  |
| $y_2$  | 0    | 0.6  | 0.1  | 0.7  | 0    | 0.5  |
| $y_3$  | 0    | 0.3  | 0.5  | 0.3  | 0.6  | 0.1  |
| $y_4$  | 0    | 0.6  | 0.1  | 0.7  | 0.6  | 0.5  |
| $y_5$  | 0    | 0.3  | 0.1  | 0.3  | 0.2  | 0.3  |
| Max Yd | 0.7  | 0.3  | 0.5  | 0.3  | 0.6  | 0.1  |

- Selecting further variables.

Since the discrimination power of indispensable variable (2.5) is less than the discrimination power of all variables (3.6), we will select more variables.

$ODP(y_2, \{y_4, y_5\}, K) = 0$, $ODP(y_4, \{y_2, y_5\}, K) = 0.4$ and $ODP(y_5, \{y_2, y_4\}, K) = 0$

Base on the *ODP* values, we select the variable *y₄* and we update *Max Yd*. We update *Max Yd* by doing the maximum, for each pair of object *(aᵢ,aⱼ)*, between the value of *g(y₄,(aᵢ,aⱼ))* and the value stored in *Max Yd* for this pair of object (see Table 3). By doing the summation of all values of Max Yd row, we calculate the discriminate power of the selected variables: $DP(({y_1, y_3, y_4}), K) = 3.6$.

Since $DP(({y_1, y_3, y_4}), K) = DP(({y_1, y_2, y_3, y_4, y_5}), K)$, we stop the algorithm. So the selected variables are: $y_1$, $y_3$ and $y_4$.

Table 3. Max Yd updated with value of the selected variable

|  | (a₁, a₂) | (a₁, a₃) | (a₁, a₄) | (a₂, a₃) | (a₂, a₄) | (a₃, a₄) |
|---|---|---|---|---|---|---|
| y₁ | 0.7 | 0 | 0.1 | 0.1 | 0 | 0.1 |
| y₂ | 0 | 0.6 | 0.1 | 0.7 | 0 | 0.5 |
| y₃ | 0 | 0.3 | 0.5 | 0.3 | 0.6 | 0.1 |
| y₄ | 0 | 0.6 | 0.1 | 0.7 | 0.6 | 0.5 |
| y₅ | 0 | 0.3 | 0.1 | 0.3 | 0.2 | 0.3 |
| Max Yd | 0.7 | 0.6 | 0.5 | 0.7 | 0.6 | 0.5 |

## 6 APPLICATION

In order to test and validate the result of our algorithm, two categories of experiments have been done. In the first experiment, we test the algorithm on real datasets, and we validated some of the results of the feature selection with an expert. The purpose of the second experiment is to do an automatic quality test. The experiment was done on a simulated data set. Base on some quality indicators we compare *Minset* and *Minset-Plus* algorithm results.

### 6.1. Real Datasets Experiment

### 6.1.1. UCI Machine Learning Repository Datasets

The datasets of this experiment come from the UCI Machine Learning Repository [16]. We used our Symbolic Object Generator program in order to create the symbolic objects which represent the clusters of the individuals of these datasets (see the section 6.2 for more details about this program). The Table 4 describes the datasets.

Table 4. Dataset description

| DataSet | Attribute Number | Individual Number | SO Number | Cluster Discrimination | % Discrimination by Extent | Type of Data |
|---|---|---|---|---|---|---|
| Iris | 4 | 150 | 3 | 100% | 48.70% | Real |
| Audiology | 69 | 226 | 24 | 100% | 100% | Categorical |
| Dermatology | 33 | 366 | 6 | 100% | 93.9% | Categorical, Integer |
| Heart Disease | 13 | 303 | 5 | 100% | 83.2% | Categorical, Integer, Real |
| Cardiotocographic | 22 | 2126 | 10 | 100% | 100% | Categorical, Integer, Real |

NOTICE: The Missing values are treated by using a substitution with a measure of central tendency.

The columns used in table 4 to describe the datasets are:

- "Attribute Number" indicates the number of attributes used the describe the symbolic objects of the dataset.
- "Individual Number" indicates the number of individuals used in the validation testing of the dataset.
- "SO Number" indicates the number of symbolic objects contained in the dataset.
- "Cluster Discrimination" indicates the percentage discrimination between the clusters. All clusters of the datasets are 100% discriminated.
- "% Discrimination by Extent" calculates the average of percentage of discrimination between objects using the selected variables. The extent is based on extent calculation on the individual data. The parameter is calculated by the following formula:

$$\% \, Discrimination \, by \, Extent = \frac{1}{|K|} \sum_{(a_i,a_j) \in K} 1 - \frac{|ext_\Omega(a_i) \cap ext_\Omega(a_j)|}{|ext_\Omega(a_i) \cup ext_\Omega(a_j)|}. \tag{16}$$

This indicator shows us also the quality of the description of the objects representing the clusters of individuals. For instance, we can notice that using the boolean symbolic objects the IRIS objects are not well discriminated.

- "Type of Data" indicates what data types are used by the variables to describe the symbolic objects.

The results of the feature selection on the cited datasets are summarized in Table 5 for *Minset-Plus* and in Table 6 for *Minset* algorithm. We can compare and analyze the results on the following parameters:

- Discrimination Power (*DP*): This parameter shows the numbers of objects couple discriminated by the selected variable. The *DP* of *Minset-Plus* is always greater than the *DP* of *Minset*, because of the use of partial discrimination. Also, we can notice that the *DP* of Minset algorithm for Iris and Heat Disease datasets is null, since all the objects are overlapped and *Minset* use Boolean discrimination to calculate the *DP*.

- % Discrimination by Extent: Comparing the values obtained by *Minset-Plus* after selection with those of datasets before the selection, we can notice that the objects after selection are still well discriminated, the average of difference between them is 1.94%. But the results of *Minset* are less good, especially when there is a big overlapping.

- % of Reduction: the result of feature selection using *Minset-Plus* was good, since the algorithm has reduced from 68% to 84%. But using Minset, the algorithm could not select any feature for IRIS and Heart Disease datasets. For Dermatology and Cardiotocographic, *Minset-Plus* selected less variables than *Minset-Plus*, but at these times *Minset* lost in the discrimination of objects after selection (see column % Discrimination by Extent).

- Time execution: the times obtained by *Minset-Plus* are very well, and when we compare them to the times obtained by *Minset-Plus*, we can see a huge difference. See the complexity test section for more detail on the complexity of the algorithm.

Table 5. Feature Selection result with Minset-Plus Algorithm

| DataSet | Discrimination Power | % Discrimination by Extent | % of Reduction | Time execution (Milliseconds) |
|---|---|---|---|---|
| Iris | 1.93 | 48.70% | 75.00% | 3 |
| Audiology | 269.74 | 100% | 84.05% | 293 |
| Dermatology | 13.91 | 94.1% | 81.81% | 45 |
| Heart Disease | 6.56 | 77.6% | 69.23% | 6 |
| Cardiotocographic | 42.20 | 100% | 68.18% | 39 |

Table 6. Feature Selection result with Minset Algorithm

| DataSet | Discrimination Power | % Discrimination by Extent | % of Reduction | Time execution (Milliseconds) |
|---|---|---|---|---|
| Iris | 00.00 | 0% | 00.00% | 7 |
| Audiology | 170.00 | 100% | 84.05% | 2346 |
| Dermatology | 11.00 | 93.9% | 87.87% | 318 |
| Heart Disease | 00.00 | 0% | 00.00% | 38 |
| Cardiotocographic | 17.00 | 95% | 86.36% | 251 |

**6.1.2. Expert Validation for Biological datasets**

The expert assessment of *Minset-Plus* result has been done on datasets in biology used by Lebbe [17] and described in the Table 7. These datasets present a light percentage of overlapping. This means that the symbolic objects of these datasets are well discriminated. The overlapping criterion is defined in (17).

$$\%overlapping = 1 - \frac{DP(Y,\ K)}{card(K)}. \tag{17}$$

Table 7. Dataset description

| DataSet | Attribute Number | SO Number | Type of Data | % overlapping |
|---|---|---|---|---|
| Tristichacees | 29 | 16 | Categorical | 0% |
| Aquatic Insects | 12 | 16 | Categorical | 0.417% |
| Siphonophores | 7 | 10 | Categorical | 0% |
| Phlebotomines | 53 | 73 | Categorical, | 0.002% |
| Felines | 17 | 30 | Categorical | 1.609% |

The Table 8 gives the names of variables selected by *Minset-Plus* and *Minset* algorithm for more details about the data see [6].

Table 8. Feature Selection result with Minset-Plus and Minset Algorithm

| Datasets | Minset-Plus feature selection | Minset feature selection |
|---|---|---|
| Tristichacees | presence of a welded part of tepals, number of stamens, net length of stamens, ovary height, bract length, distribution, type | Tepal length, presence of a welded part of tepals, , number of stamens, net length of stamens, ovary height, distribution, distribution in Asia, type |

| Aquatic insects | extremity of the abdomen, articulated legs free, mouthparts, type of lateral abdominal bronchia, lifestyle, prolegs, extremity of prolegs, development stage | extremity of the abdomen, articulated legs free, mouthparts, type of lateral abdominal bronchia, lifestyle, extremity of prolegs |
|---|---|---|
| Siphonophores | number of crests, stomatocyte, apex, pigment, form of crests, type | number of crests, stomatocyte, apex, pigment, form of crests, type |
| Phlebotomines | Total length of genital pump, Course of the terminal portion, Length of genital filaments, Ratio filament / pump, Length of lateral lobe, Shape of Paramere, Number of setae on Coxite tuft, Distribution of 4 style spines, Distribution of 5 style spines, Total length of PALPS, Ratio P5/P3, Ratio P3/P4, Proximal prolongation of Ascoids, Distal prolongation of ASCOIDS, Length of F1, Ratio (Total Palp)/F1 | Total length of genital pump, Course of the terminal portion, Striation of genital filaments, Length of genital filaments, Ratio filament/pump, Length of lateral lobe, Shape of Paramere, Number of setae on Coxite tuft, Distribution of 4 style spines, Distribution of 5 style spines, Palpal formula, Total length of PALPS, Ratio P3/P4, Distal prolongation of Ascoids, Length of F2, Ratio (Total PALP)/F1, Groups of Lutzomyia |
| Felines | appearance of the coat, hair length, relative length of the tail respect to the body, hours of predator behavior, prey size, continent | appearance of the coat, hair length , hours of predator behavior, continent, height at the withers, main method of hunting |

Using the datasets described in Table 7, we noticed that *Minset* and *Minset-Plus* most of the time obtained the same result (see Fig. 3). The expert has validated and accepted the variables selected by *Minset-plus*.

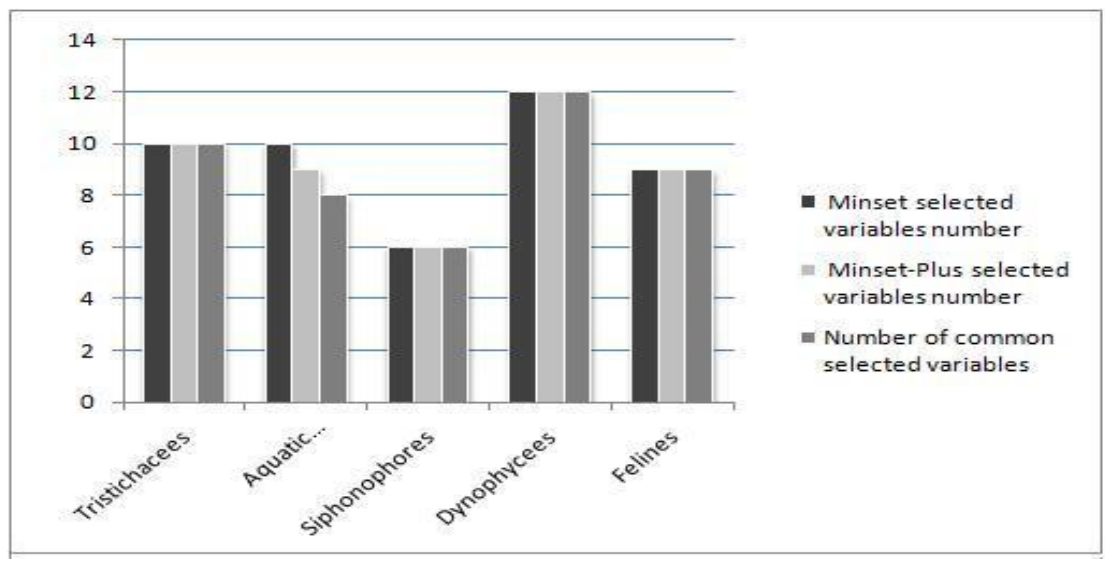

Figure 3. Comparison of number of selected variables

On the dataset of "aquatic insects" and "*Phlebotomines*", *Minset-Plus* selected one variable less than *Minset*, however the expert has assess that the variables selected by *Minset-Plus* discriminate better the objects than the variables selected by *Minset*, since all indispensables variables selected by *Minset* have been also selected by *Minset-Plus*, and the variables which were selected by *Minset-Plus* and not selected by *Minset* have a good add-value in discrimination.

**6.2. Automatic Quality Test**

In order to provide an automatic quality test, we designed and developed a complete system (see Fig. 4).

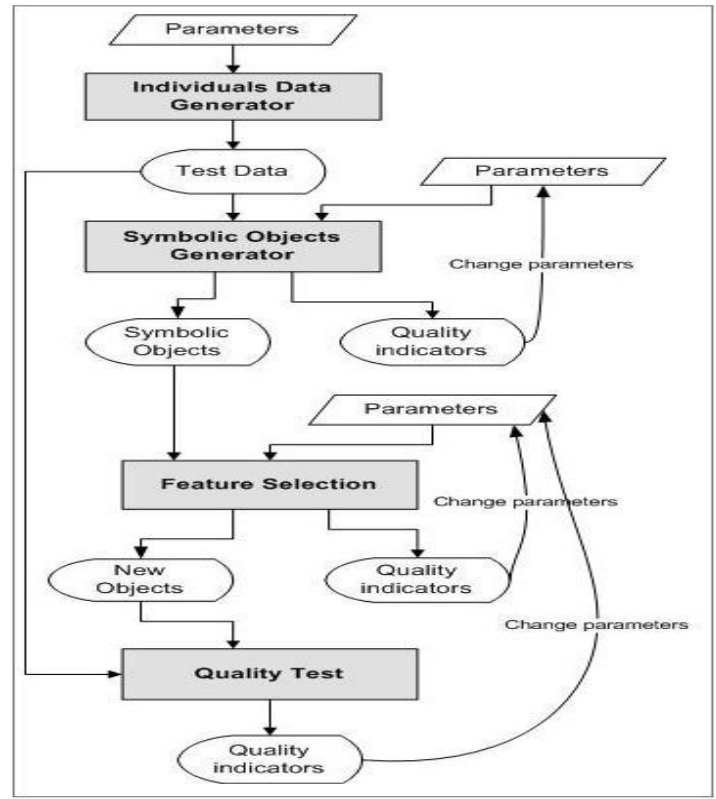

Figure. 4. Automatic Quality Test Loop

This system includes the following modules:

- Individuals Data Generator: When the datasets of symbolic objects don't have a set of individuals for testing, we utilize the Individuals Data Generator module to generate individual data. This module uses the domains of the descriptive variables, and the values taken by variables in the symbolic objects to generate the individuals.

- Symbolic Objects Generator: It is a module used to generated symbolic objects from given individual set. Each individual should be associated to one or many clusters. We provide for each variable an input parameter which indicates what type of variable we will have as output in the description of the generated symbolic objects (Boolean, categorical, set of numerical values, interval). To generate a set of values, the generator used the union operator to create the output set. For the interval, the generator will use the minimum and the maximum value of individuals belong to the cluster represented by the generated symbolic objects. The generator also takes into consideration the domain of the variables and values taken by the individuals of other clusters, to refine the generated interval. For instance:

Let have four individuals described by one variable height. The first two individuals belong to the cluster of small and tall people and the two last individuals belong to the average tall people. $I_1=(150), I_2=(190), I_3=(165)$ and $I_4=(170)$

If we take into consideration only the minimum and maximum values, the symbolic object generator will generate these two symbolic objects:
$a_1$=[ height=[150 190] ] and $a_2$=[ height=[165 170] ]

We can notice that the object $a_1$ is over generalized and it extend will include the individuals of the average tall people cluster, this is why the generated interval of object $a_1$ will be refined by taking into consideration the values of individuals of the average tall people cluster: $a_1$=[ height=[150 165[ ]170 190] ]

- Feature selection: To compare the results, we can use the following three algorithms: *Minset*, *Minset* using partial discrimination for feature selection criterion, and *Minset-Plus*. We have as output new symbolic objects, described with the selected variables, and we have also some quality indicators such as: the reached discrimination power, the percentage of reduction and the time of execution.
- Quality test: this module tests the quality of the new symbolic objects (objects using the selected variables). We get as output some quality indicators such as: real object extents and percentage of object overlapping. Base on these indicators we can change some parameters to improve the feature selection.

Using the described system, we did two categories of tests: Quality test and complexity test:

### 6.2.1. Quality Test

Assessment of the number of selected variables: we generated fifteen symbolic object datasets with overlapping percentages sliding from 0.10% to 16%. You can notice in Fig. 5, that when the overlapping percentage is low, *Minset* and *Minset-Plus* algorithm select the same number of variables. However *Minset* algorithm selects less variables when we increase the percentage of overlapping. Also, the number of the selected variables decreases dramatically with the increase of the overlapping percentage. When we reached 16% of overlapping percentage, no variable has been selected. In the other side, the number of selected variables of *Minset-Plus* algorithm decreases slightly; and sometimes it is stationary for several percentages of overlapping. This proves that the selection criterion used by *Minset-Plus* algorithm is more adequate than the one use by *Minset* algorithm, if we are dealing with overlapping clusters.

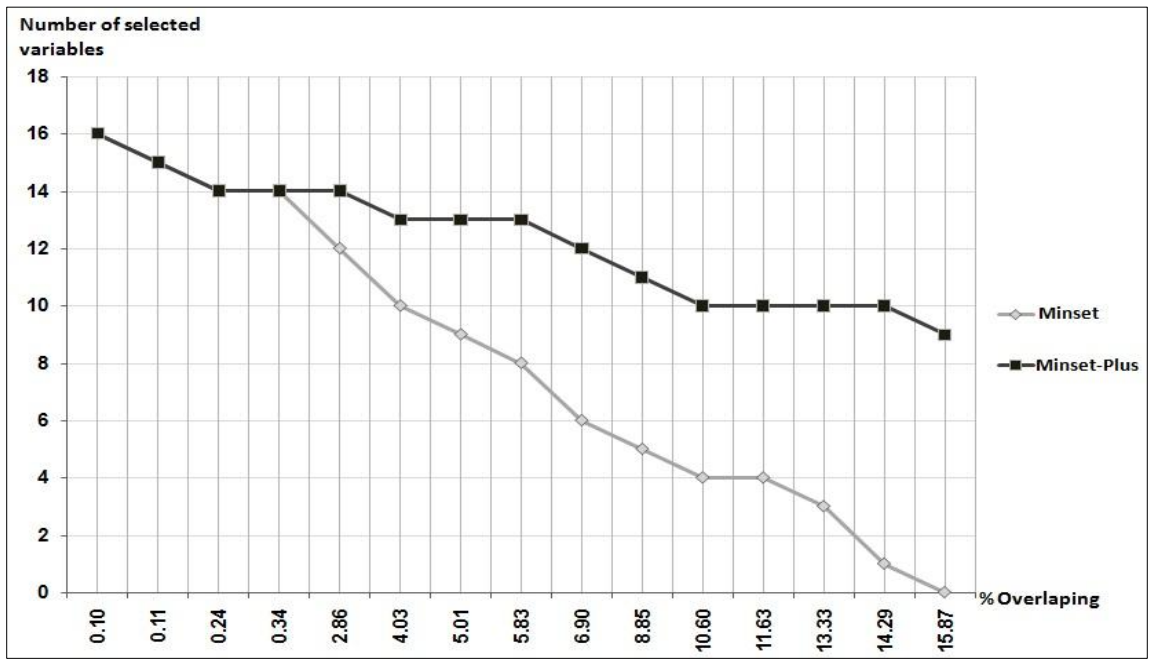

Figure 5. Number of selected variables vs. % of overlapping

To understand why *Minset* algorithm selects fewer variables than *Minset-Plus* algorithm, we calculated the discriminant power (*DP*) for the selected variables. You can notice in Fig. 6 that the *DP* of variables selected by *Minset* is too much low than the *DP* of variables selected by *Minset-Plus*. This is due to the fact that *Minset* use the boolean discrimination in the selection criterion. For example, if two clusters have an intersection in only one individual among 1000 individuals, *Minset* considers that these two clusters are overlapped, and they cannot be discriminate by any variable. However, in the same situation *Minset-Plus* selects the best variables which discriminate the 999 left individuals. You can also see in Fig. 5 that around 14% of overlapping *Minset-Plus* has selected 10 variables and *Minset* selected only 1. But, in Fig. 6 you can notice that the *DP* reached at this level by the selected variables is: 690 for *Minset-Plus* and 3 for *Minset*.

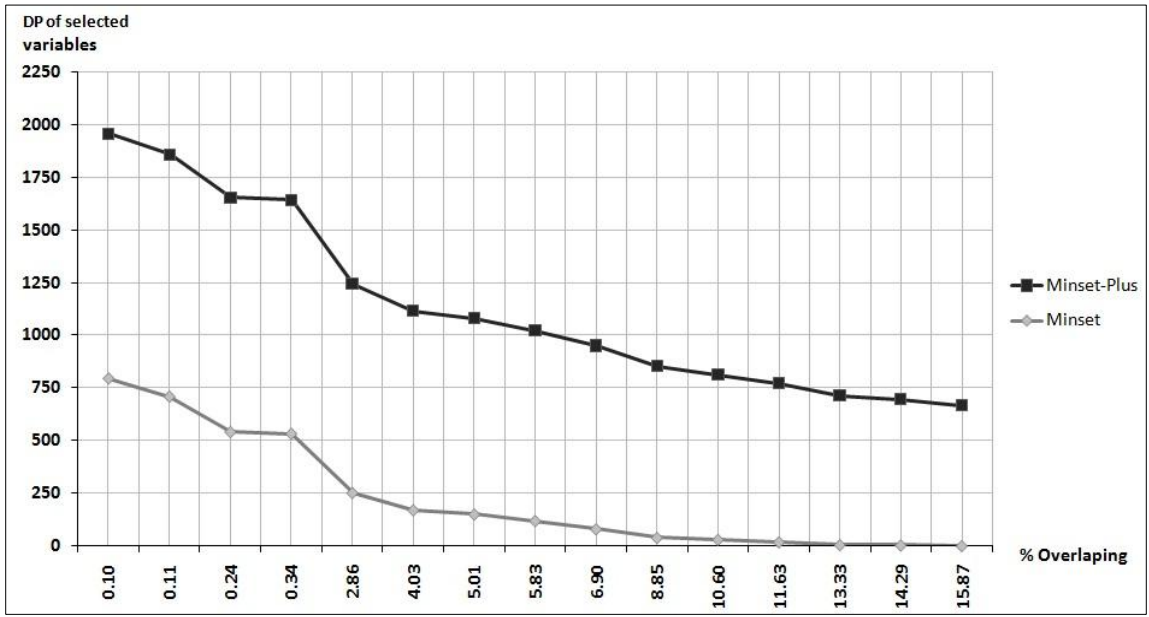

Figure 6. Discriminant Power of Minset and Minset-Plus

During the execution of the feature selection with *Minset-Plus*, we noticed that with only few selected variables the algorithm can reach between 95% and 99% of percentage of discrimination, and a big number of variables are selected in order to reach 100% of discrimination. So base on this remark, we made a new experiment, where we compare the number of selected variables of *Minset-Plus* at 99% of discrimination and the number of selected variables of *Minset-Plus* and *Minset* at 100% of discrimination (see Fig. 7).

The first finding of this experiment is that with 99% of discrimination, the number of selected variables with *Minset-Plus* is too much low than the number of selected variables with *Minset-Plus* and *Minset* with 100% of discrimination. In Fig. 7, we can see a gap of 10 variables between *Minset-Plus 99%* and *Minset-Plus*, and a gap of 6 variables between *Minset-Plus 99%* and *Minset*. We can notice also that the number of selected variables with *Minset* with 99% of discrimination is nearly the same as *Minset* with 100% of discrimination. The second finding of this experiment is: with the increasing of the overlapping percentage, *Minset-Plus* with 99% of discrimination selects mostly the same number of variables as *Minset-Plus*. we can see also in Fig. 7 that from 3% to 14% of discrimination, the number of selected is always 4. This can be explained be the following: *Minset-Plus* selects the variables using a selection criterion base on partial discrimination, this selection criterion allows to select the best variables which maximize the discriminate between the symbolic objects. So we can conclude that the selection criterion with partial discrimination is too much better than the selection criterion with boolean discrimination, especially when the overlapping is high.

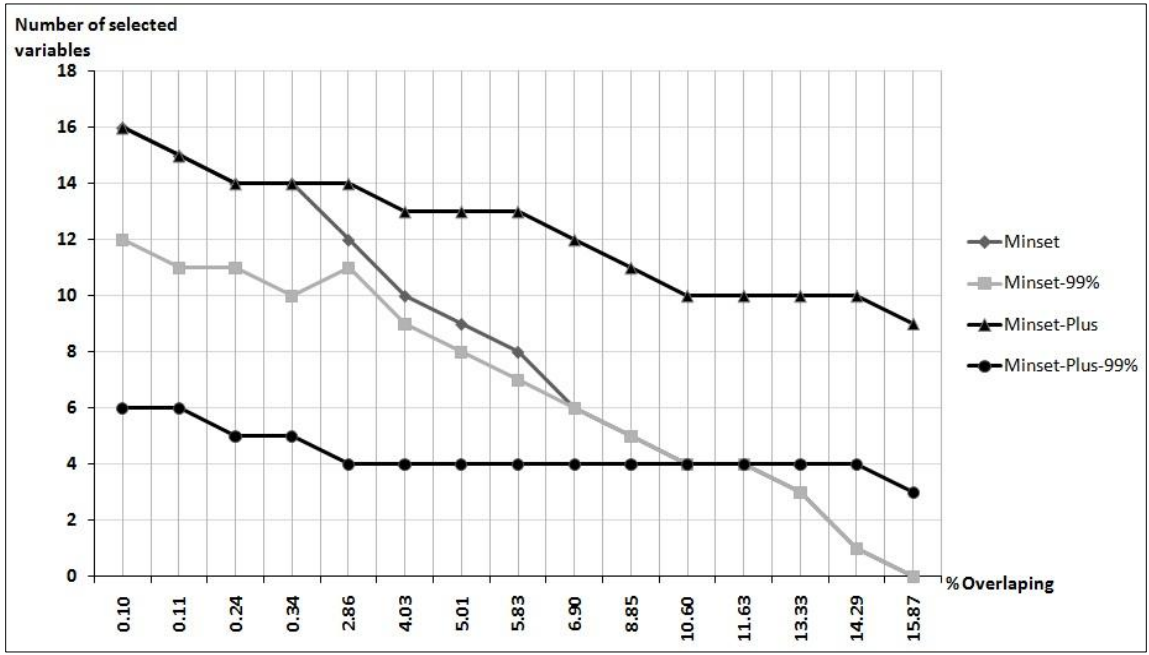

Figure 7. Extent intersection average of the new objects

Another category of experiment has been done to assess the quality of the selected variables. In this experiment, we used 300 individuals to generated 10 datasets. The datasets are described by 20 variables. As shown in Fig. 8, the number of objects used by the datasets varies from 10 to 100. After doing the feature selection using *Minset* and *Minset-Plus* algorithms, we generated new symbolic objects described by the selected variables. We calculated the real intersection extent average of these new objects on the 300 individuals. Finally, we compared these results with the real intersection extent average of the symbolic objects described with all variables (before selection). The purpose of this comparison is to know if the selected variables will discriminate correctly the clusters described by the symbolic objects. So a good feature selection should minimize this difference.

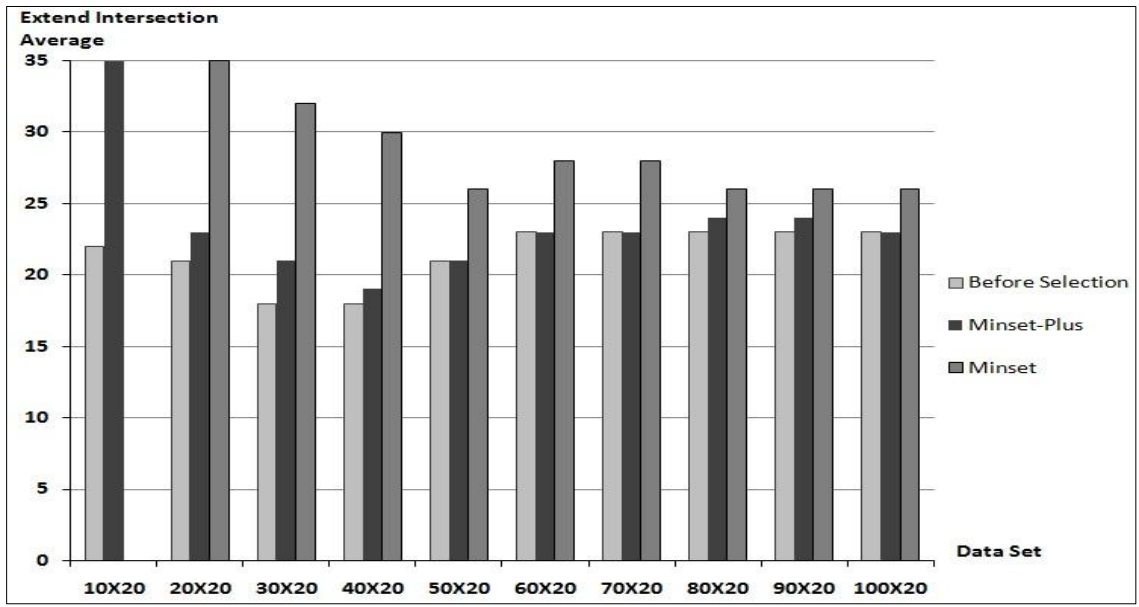

Figure 8. Extent intersection average of the new objects

We can notice that the selected variables by *Minset-Plus* are too much better that those selected by *Minset*. Since the difference between the real intersection extent average of object used by *Minset-Plus* before and after selection (described only by the selected variables) is low, comparing to the corresponding value of *Minset* (we notice that in the dataset 10×20 *Minset* did not select any variable because of all variable is equal to zero).

**6.2.2. Complexity Test**

We executed feature selection on the 10 datasets (same datasets used in the test of Fig. 8). We compared the time execution of *Minset-Plus* and the time execution of *Minset Partial* (it is the algorithm *Minset*, but using the partial discrimination as selection criterion. This means this algorithm did not use the mathematical properties and the discrimination matrix used by Minset-Plus to reduce the complexity). This test shows that the mathematically properties and discrimination matrix used in *Minset-Plus* algorithm have an impact in reducing the complexity of the algorithm.

The selection criterion based on partial discrimination uses complex operations, so this is why we noticed that the complexity grow exponentially when we increase the number of symbolic objects (see the time execution of *Minset Partial* in Fig. 9). In meantime, the time execution of *Minset-Plus* has grown slowly when we compare it to the time execution of *Minset Partial*.

In the last experiment, we compared the complexity of *Minset-Plus* algorithm using the feature selection criterion based on partial discrimination, and *Minset* algorithm using feature selection criterion based on Boolean discrimination. We executed the feature selection on the same datasets used in the experiment of Fig. 9. We can notice in Fig. 10 that even when we compare *Minset-Plus* complexity with *Minset* complexity; *Minset-Plus* is faster than *Minset*, and the time execution of *Minset-Plus* grow slowly than the time execution of *Minset*.

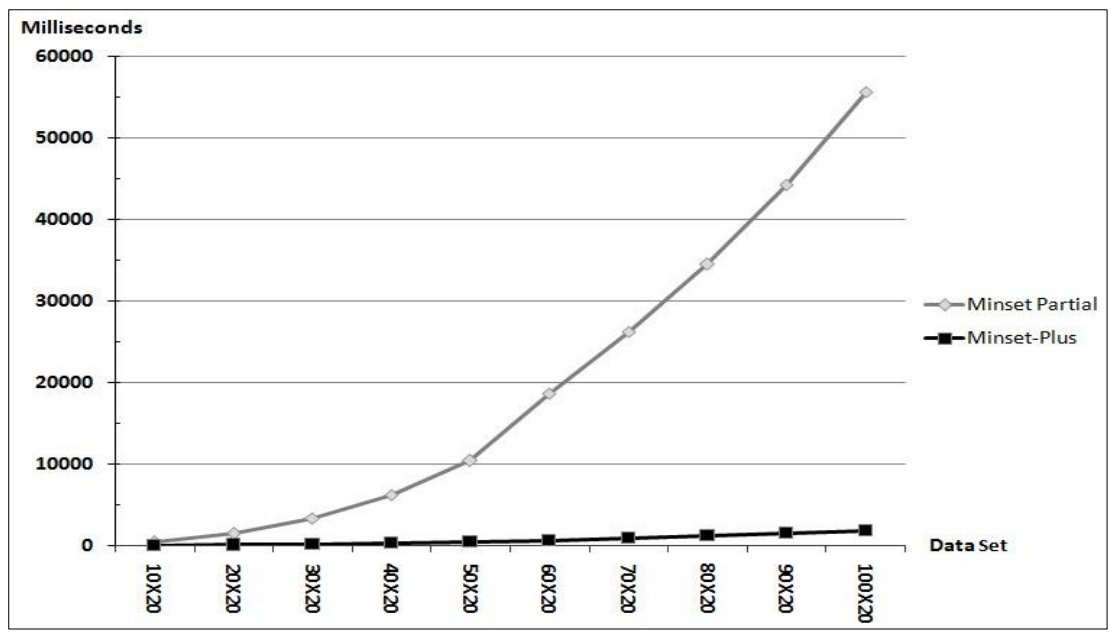

Figure 9. Complexity of Minset-Plus vs. Minset Partial

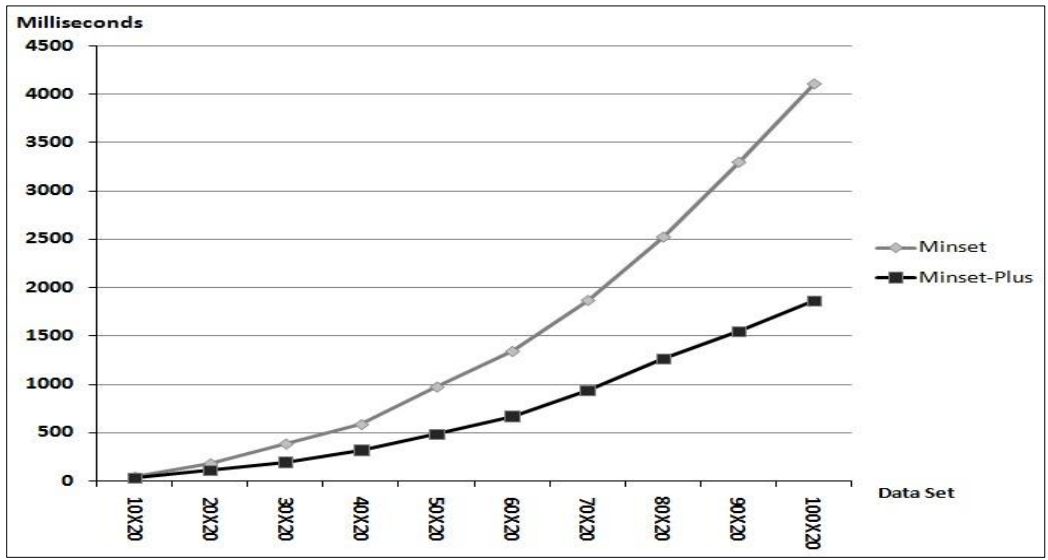

Figure 10. Complexity of Minset-Plus vs. Minset

## 7. CONCLUSIONS

In this paper, we have presented the algorithm *Minset-Plus* which is an extension and improvement of the algorithm *Minset*. *Minset-Plus* is an algorithm for selecting discrimination variables on a set of symbolic objects. The use of partial discrimination allows to process deeply and in a better way the selection of the variables. Based on the discovery of some mathematical properties on *ODP* and *DP* functions, and by the use of the discrimination matrix, we obtained a huge improvement on the complexity of the algorithm.

The quality of the selected variables is based tightly on the function *"g"* defined in (10) and on some parameters that the expert can introduce in the algorithm to refine the selection of variables. Therefore, to appraise the robustness of this function and to measure the quality of the selected variables, we will introduce in our next research automatic validation using expert inputs.

Also, we plan to integrate *Minset-Plus* algorithm with databases, and this could be done by implementing in the algorithm as a plug-in in a database in order to access the data sets in an efficient and faster way.

## ACKNOWLEDGEMENTS

The authors would like to thank King Saud University, the college of Computer and Information Sciences, and the Research Center for their sponsorship.

**Authors**

Djamal Ziani is Professor Assistant in King Saud University in Computer Sciences and Information Systems College from 2009 until now. Researcher in data management group of CCIS King Saud University. He received Master degree in Computer Sciences from University of Valenciennes France in 1992, and PH.D. in Computer Sciences from University of Paris Dauphine, France in 1996.

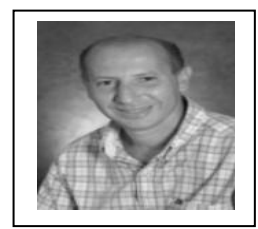